# Lattice constant variation and complex formation in Zincblende Gallium Manganese Arsenide

G. M. Schott, W. Faschinger and L. W. Molenkamp

Physikalisches Institut, Universität Würzburg, Am Hubland, D-97074 Würzburg, Germany

Abstract

We perform high resolution X-ray diffraction on GaMnAs mixed crystals as well as on GaMnAs/GaAs and GaAs/MnAs superlattices for samples grown by low temperature molecular beam epitaxy under different growth conditions. Although all samples are of high crystalline quality and show narrow rocking curve widths and pronounced finite thickness fringes, the lattice constant variation with increasing manganese concentration depends strongly on the growth conditions: For samples grown at substrate temperatures of 220°C and 270°C the extrapolated relaxed lattice constant of Zincblende MnAs is 0.590 nm and 0.598 nm respectively. This is in contrast to low temperature GaAs, for which the lattice constant decreases with increasing substrate temperature.

In the last years the new semimagnetic semiconductor Gallium Manganese Arsenide [1] (GaMnAs) has attracted considerable attention, especially in the context of "spintronics" or the manipulation of spin in traditional semiconductors. Compared to other semimagnetic semiconductors, GaMnAs has a number of peculiarities: Mn acts as an acceptor in GaAs, so that the hole concentration in GaMnAs is typically very high. The free holes influence the coupling between Mn spins, making GaMnAs ferromagnetic with a Curie temperature which depends on the hole concentration, and theory predicts even ferromagnetism at room temperatures for very high hole concentrations [2]. In practice, however, the obtained hole concentrations are much lower than the concentration of substitutional Mn. This is a consequence of the fact that the solubility of Mn in GaAs is very low at thermal equilibrium, so that GaMnAs is a metastable compound which can only be grown with low temperature MBE far from thermal equilibrium. However, in GaAs a low growth temperature leads to a very high concentration of As antisites [3] which are deep donors that can compensate the holes in GaMnAs. A proper understanding of point defect formation in GaMnAs is thus important, since the resulting compensation directly influences the Curie temperatures [4,5].

The high antisite concentrations in low temperature GaAs lead to an increase in the lattice constant [3], which can be precisely measured with high resolution X-ray diffraction (HRXD). This increase depends strongly on the growth conditions and can be as large as 0.01 nm (or about 0.2% of the bulk GaAs lattice constant) for high As flux and low substrate temperature. Similar growth conditions are usual in GaMnAs MBE: For example, the lattice constants of GaMnAs given by Ohno et al. extrapolate to a GaAs lattice constant of 0.566 nm for zero Mn content [6] instead to the bulk value of 0.565325 nm. In addition, there are experimental indications that the As antisite concentration increases with the incorporation of Mn [7], so that even stronger effects of the growth conditions on the lattice constant can be expected for GaMnAs.

As a consequence, we performed a systematic HRXD investigation on GaMnAs samples grown by MBE at different conditions. The first rationale was to reduce the number of As antisites in the GaAs starting material. We thus investigated the GaAs lattice constant as a function of substrate temperature and As to Ga beam pressure ratio. At a substrate temperature ($T_{Sub}$) of 270°C and a beam pressure ratio of 5 there is no measureable influence of the antisites on the GaAs lattice constant. A first series of GaMnAs samples with Mn source temperatures from 640°C to 670°C was grown at these conditions. The RHEED pattern indicated polycrystalline growth for Mn source temperatures higher than 670°C, so that the Mn content in this series is relatively low. A second GaMnAs series was then grown at $T_{Sub}$= 220°C and identical beam pressure ratio for Mn source temperatures between 640°C and 710°C. At these conditions the lattice constant for GaAs measured by HRXD is 0.5656 nm, indicating that the degree of As antisite incorporation is similar to the growth conditions given in Ref. [6]. The HRXD measurements were performed with a Philips Xpert diffractometer equipped with a copper tube, a parabolic multilayer mirror, a four crystal monochromator and a double crystal analyzer in front of the detector. Independent of the growth conditions, all samples presented in this study are of high crystalline quality with rocking curve widths comparable to those of the GaAs substrate and show pronounced finite thicknes fringes indicating flat interfaces and surfaces. As previously reported for similar structures [6, 8], all layers are fully pseudomorphic with respect to the GaAs substrate, as verified by reciprocal space maps around the asymmetric 115 reflex. As in Ref. [6], all lattice constants that are given in the following are relaxed values which were calculated from the measured vertical lattice constants of the tetragonally distorted layers under the assumption that the elastic constants of GaMnAs are equal to those of GaAs.

If the As antisite incorporation and thus the GaMnAs lattice constant indeed depends on the Mn content, one cannot simply determine the Mn content x by comparing the vertical lattice constant to literature data, as is usually done. Therefore we grew a 2 μm thick calibration layer with a high Mn content at $T_{Sub}$= 220°C and measured the Mn content with energy dispersive analysis of X-rays (EDX). The lattice constant of 0.5669 nm of this sample would correspond to a Mn content of about 4% according to the calibration given by ref. [6]. However, the measured EDX value is about 50% larger, namely 6.2%. Nevertheless, without the use of a proper calibration sample, the error of the EDX measurement is relatively large. In order to measure the Mn content more precisely with an independent method, we grew a special calibration sample at identical growth conditions consisting of a GaMnAs/GaAs superlattice (SL2) with thick GaMnAs and thin GaAs layers on top of an inverted superlattice (SL1) with thin GaMnAs and thick GaAs layers. Both SLs have an intended period of about 10 nm. The HRXD spectrum of this sample around the 004 Bragg refection is shown in Fig.1. Besides the central SL peaks, clear and very sharp satellite reflections are visible. The thin line in Fig.1 is a fit based on dynamical scattering theory, which allows a very precise determination of the two SL periods. The accuracy of the period is better than 0.02 nm, as demonstrated by the insert in Fig.1, which shows a zoom up of one satellite together with a perfect fit and a fit with a SL period which is 0.02 nm larger (labeled +0.02 nm). From these periods and the known growth times for each layer the GaAs and GaMnAs growth rates, and consequently the Mn content x, can be determined self-consistently under the assumption that all fluxes were constant during growth. This assumption was veryfied by growing 300 nm thick GaMnAs calibration samples before and after the SL growth which had nearly identical thickness and composition. For the given structure the resulting Mn content is 6.0±0.5%, which is in good agreement with the EDX measurement. Consequently, the extrapolated lattice constant of pure Zincblende MnAs resulting from our calibration sample is only 0.590±0.02nm compared to a value of 0.598 nm given in ref. [6]. This seems to indicate that the point defect incorporation is indeed influenced by Mn incorporation, substrate temperature and As flux.

However, before such a conclusion can be really drawn, one must ensure that in the calibration sample, which has a relatively high Mn content, no phase separation occurs, as this would probably make some of the Mn invisible for the X-ray measurement. In the narrow temperature range in which the Mn source was operated, the Mn evaporation rate and thus the Mn flux is expected to be proportional to $\exp(-E_a/kT)$, where $E_a$ is an activation energy and k is the Boltzmann constant. If all Mn is incorporated in the same way, without any phase separation, the observed change of the lattice constant should be proportional to the Mn flux. In order to test this assumption, the insert in fig.2 shows the change of the vertical GaMnAs lattice constant (relative to the vertical lattice constant of low temperature GaAs) for both sample series is plotted as a function of the inverse Mn source temperature. Within the error of the experiment, both data sets indeed lie on straight lines, so that significant phase separation can be excluded. Nevetheless the slopes of the two lines are different. This means that the increase in lattice constant is not a function of the Mn flux alone (which is identical for both sample sets), but additionally depends on the incorporation of something else.

To make this still more evident, in Fig.2 the lattice constant of both sample series is plotted as a function of their Mn content (circles). The SL calibration sample (solid square) and the EDX calibration sample (solid diamond) are also included. Since a reliable calibration is only possible for high Mn content, all calibrations were done for samples grown at 220°C. The data points of the series at $T_{Sub}$= 270° (open circles) are included in the plot under the assumption that their total Mn content is identical to that of the corresponding 220°C-samples grown under identical flux conditions. We think that this assumption is reasonable since these sample pairs were always grown in the same growth run and the sticking coefficient of Mn is most probably unity at the low substrate temperatures used. It is evident that the lattice constant increase with x is very different for the two substrate temperatures used. Unfortunately we could not grow GaMnAs samples at $T_{Sub}$= 270°C with a high enough Mn content to verify wether this trend is continued for higher x. However, it has been shown recently that superlattices containing sub-monolayer coverages of pure Zincblende MnAs can be grown at substrate temperatures as high as 280°C [9]. We grew such SLs at $T_{Sub}$= 270°C and the source conditions used for the GaMnAs samples with a Mn content of 2%. Since the thickness and Mn content of these samples can be precisely measured, the Mn flux at these conditions and consequently the MnAs coverage in a SL can be calculated. Fig. 3 shows a 004 HRXD spectrum of such a SL with a MnAs coverage of 0.18 monolayers per period. Since the coverage is well known, the position of the central SL peak only depends on the vertical MnAs lattice constant, which can be precisely determined from a fit of the diffraction pattern. Such a fit for a perfectly abrupt SL is shown as the thin line in Fig.3. Real superlattices of that type show significant interdiffusion [9]. However, since X-ray diffraction averages over the given amount of MnAs lattice cells, independent of their actual vertical or lateral position in the SL, these imperfections do not influence the evaluation of the MnAs lattice constant, but only the relative satellite intensities. The resulting MnAs lattice constant is 0.598±0.02 nm, with the error mainly determined by the uncertainty in the Mn content of the calibration samples. This value corresponds to the value given in ref. [6] and is indeed larger than the value at $T_{Sub}$= 220° C. The dashed line in Fig.3 is a linear interpolation between the GaAs lattice constant and this value and confirms that there is indeed a crossover between the 220°C- and 270°C data at higher x. For completeness, the extrapolation given by ref. [6] is also included in Fig.3 as a dotted line.

Two conclusions can immediately drawn from Fig. 2: First, since the growth conditions strongly influence the MnAs lattice constant, a determination of the Mn content based on literature calibration values can lead to significant errors. This, in turn, can be of importance for the accuracy of theories which try to determine the Curie Temperature as a function of x. Second, and probably more important, the incorporation of As for a given As flux, which is the only factor that can produce a lattice constant increase, obviously depends on both substrate temperature and Mn content: While at low x the As antisite formation is suppressed at $T_{Sub}$= 270°C, the situation at high Mn content is reversed. The simplest model to understand such a behavior is the assumption that at low substrate temperature the As antisites and the Mn ions are uncorrelated, while at high substrate temperatures the formation of a defect complex consisting of Mn and excess As becomes possible. In order to explain the observed huge variations in lattice constants, the concentration of such complexes must be on the order of some $10^{20}cm^{-3}$, and obviously the understanding and eventually the control of such enormous defect concentrations should be an important goal in GaMnAs epitaxy.

In conclusion, we have shown that the lattice constants of GaMnAs and MnAs depend strongly on the growth conditions. In contrast to low temperature GaAs, whose lattice constant increases with decreasing substrate temperature at constant beam fluxes due to additional As incorporation, the resulting MnAs lattice constant decreases by 1.4% when the substrate temperature is lowered from 270°C to 220°C. This result shows that the As incorporation strongly depends on both substrate temperature and Mn content, presumably due to the formation of Mn-As-clusters at higher growth temperatures.

Figure Captions

Fig.1: (004) high resolution X-ray diffraction pattern of a calibration sample consisting of two superlattices with thin GaMnAs layers (SL1) and thick GaMnAs layers (SL2). The insert shows the -1-order satellite of SL2 together with a perfect fit (dotted) and a fit with a period which is 0.02 nm larger (dashed)

Fig.2: Vertical lattice constant of GaMnAs layers grown at different substrate temperatures as a function of the manganese content (circles). The solid line is a linear extrapolation between two calibration points (solid squares), the dashed line a linear extrapolation between the GaAs bulk lattice constant and the MnAs lattice constant determined from a GaAs/MnAs superlattice. The dotted line is calculated from data given in ref. 6. The insert shows the change of the vertical lattice constant for GaMnAs layers grown at two substrate temperatures as a function of the manganese source temperature.

Fig.3: (004) high resolution X-ray diffraction pattern of a GaAs/MnAs superlattice grown at a substrate temperature of 270°C together with a simulation. The insert shows the central superlattice peak together with a simulation for the MnAs lattice constant of 0.612 nm (dashed) and 0.628 nm (dotted)

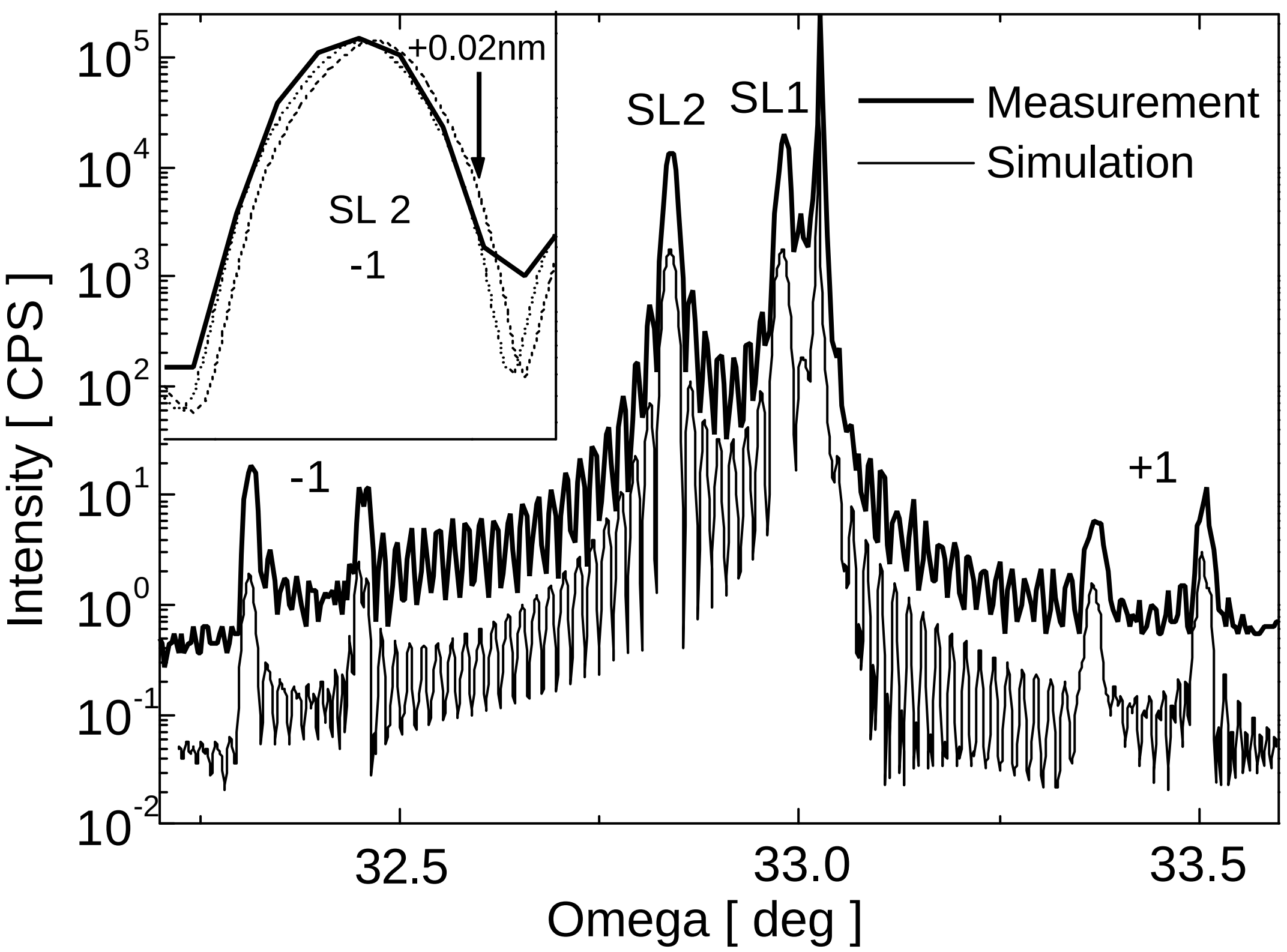

Fig.1

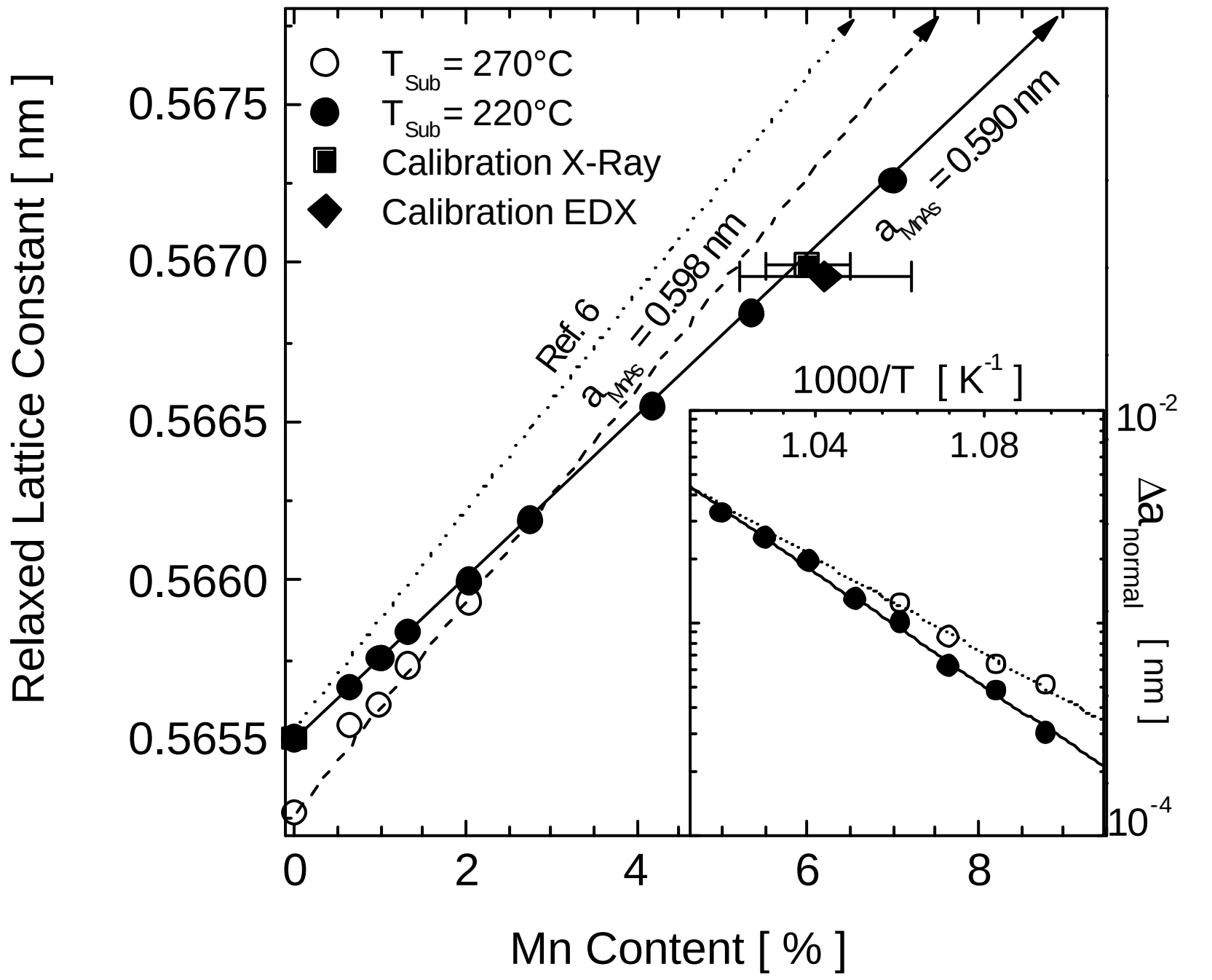

Schott et al., Fig.2

Fig.2

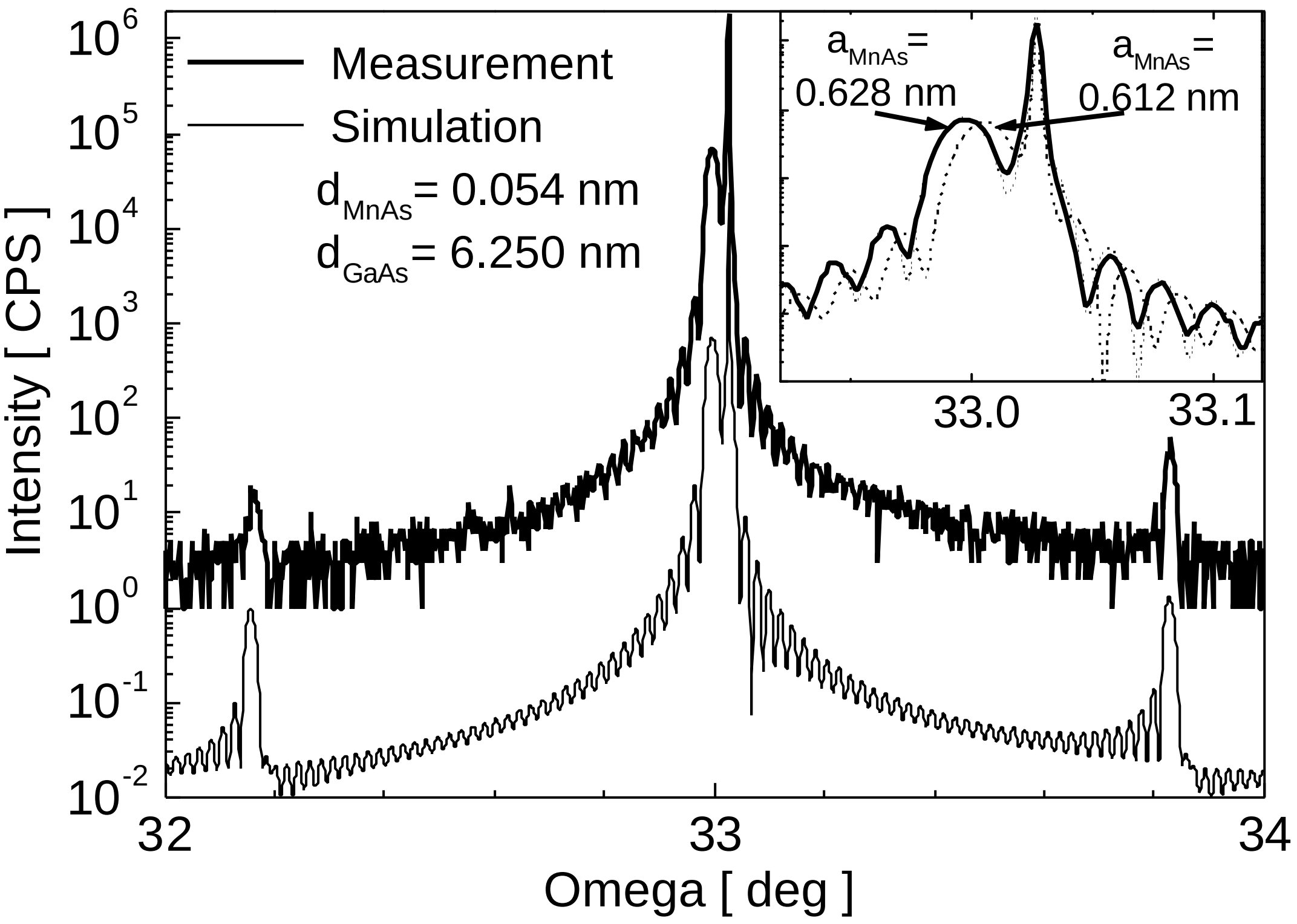

Fig.3